\documentclass{ws-ijmpcs}

\usepackage{amsmath,graphicx,verbatim,epsfig}
\usepackage{epstopdf}
\usepackage[utf8x]{inputenc}
\usepackage{amsmath,graphicx}
\usepackage[normalem]{ulem}
\usepackage{datetime}
\usepackage[colorlinks=true,linkcolor=blue,citecolor=blue]{hyperref}
\usepackage{breakurl}
\usepackage{slashed}

\newcommand{\nn}{\nonumber}

\newcommand{\be}{\begin{equation}}
\newcommand{\ee}{\end{equation}}
\newcommand{\bea}{\begin{eqnarray}}
\newcommand{\eea}{\end{eqnarray}}
\newcommand{\balign}{\begin{align}}
\newcommand{\ealign}{\end{align}}

\newcommand{\bg}{\begin{gather}}
\newcommand{\foma}{\end{gather}}

\newcommand{\noopsort}[1]{}

\def\<{\langle}
\def\>{\rangle}

\def\g{\gamma}

\def\({\left(}
\def\[{\left[}
\def\){\right)}
\def\]{\right]}

\def\sin{\hbox{sin}}
\def\ln{\hbox{ln}}

\newcommand{\ben}{\begin{eqnarray}}
\newcommand{\een}{\end{eqnarray}}
\newcommand{\nnu}{\nonumber\\}
\newcommand{\bef}{\begin{figure}[htb]\centering}
\newcommand{\eef}{\end{figure}}

\def\trimmarks{}

\begin{document}

\markboth{Echevarria, Idilbi, Kang, Vitev}
{Sivers Asymmetry with QCD Evolution}

%
\catchline{}{}{}{}{}
%

\title{Sivers Asymmetry with QCD Evolution}

\author{MIGUEL G. ECHEVARRIA\footnote{
Speaker.}}
\address{Nikhef and Department of Physics and Astronomy, 
VU University Amsterdam, \\
Science Park 105, NL-1098 XG Amsterdam, the Netherlands
\\
m.g.echevarria@nikhef.nl}

\author{AHMAD IDILBI}
\address{Department of Physics, 
Pennsylvania State University, University Park, PA 16802, USA
\\
aui13@psu.edu}

\author{ZHONG-BO KANG}
\address{Theoretical Division, 
Los Alamos National Laboratory, Los Alamos, NM 87545, USA
\\
zkang@lanl.gov}

\author{IVAN VITEV}
\address{Los Alamos National Laboratory, Los Alamos, NM 87545, USA
\\
ivitev@lanl.gov}

\maketitle


\begin{abstract}

We analyze the Sivers asymmetry in both Drell-Yan (DY) production and semi-inclusive deep inelastic scattering (SIDIS), while considering properly defined transverse momentum dependent parton distribution and fragmentation functions and their QCD evolution.
After finding a universal non-perturbative spin-independent Sudakov factor that can describe reasonably well the world's data of SIDIS, DY lepton pair and $W/Z$ production in unpolarized scatterings, we perform a global fitting of all the experimental data on the Sivers asymmetry in SIDIS from HERMES, COMPASS and Jefferson Lab.
Then we make predictions for the asymmetry in DY lepton pair and $W$ boson production, which could be compared to the future experimental data in order to test the sign change of the Sivers function.


\end{abstract}


\section{Introduction}	

Transverse spin physics has become a very active field of research both experimentally and theoretically, providing valuable information on the hadron substructure.
This information is encoded in the so-called transverse momentum dependent parton distribution and fragmentation functions (TMDPDF/TMDPFF)\cite{collins-book,Echevarria:2012js}.
One of the most analyzed spin asymmetries is the Sivers effect, originated from a particular TMDPDF called Sivers function~\cite{Sivers:1989cc}, which reresents the distribution of an unpolarized quark inside a transversely polarized hadron.
This function is not exactly universal, but has a time-reversal modified universality~\cite{Brodsky:2002cx,Collins:2002kn,Boer:2003cm,Kang:2011hk,Kang:2009bp}, which means that the Sivers function in semi-inclusive deep inelastic scattering (SIDIS) and in the Drell-Yan (DY) process are equal in magnitude but opposite in sign.
In this proceedings we review our recent phenomenological extraction of Sivers function~\cite{Echevarria:2014xaa} from current SIDIS data and the consequent predictions for DY processes, which could be compared with future experimental data.

\section{QCD evolution of TMDs}

The energy evolution of a generic TMD $F(x, b; Q)$~\cite{collins-book,Echevarria:2012js} in the impact parameter space is given by~\cite{collins-book,Aybat:2011zv,Aybat:2011ge,Echevarria:2012pw}:
\begin{align}
F(x, b; Q_f) &= F(x, b; Q_i) \exp\left\{-\int_{Q_i}^{Q_f} \frac{d\mu}{\mu} 
\left(\Gamma_{\rm cusp}\ln\frac{Q_f^2}{\mu^2}+\gamma^V\right)\right\}
\left(\frac{Q_f^2}{Q_i^2}\right)^{-D(b;Q_i)}
\,,
\nn\\
\frac{dD}{d\ln\mu} &= \Gamma_{\rm cusp}
\,,
\label{evolution}
\end{align}
where $\Gamma_{\rm cusp}$ and $\g^V$ represent the cusp and non-cusp anomalous dimensions~\cite{Echevarria:2012pw}, respectively, and $Q_i$ and $Q_f$ the initial and final scales.
The evolution kernel is valid only in the perturbative region $1/b\gg \Lambda_{\rm QCD}$. 
The function $F(x, b; Q)$ represents any TMD, in particular the unpolarized TMDPDF and TMDPFF and the $k_\perp$-weighted Sivers function:
\begin{align}
f_{q/A}(x, b; Q) &= \int d^2k_\perp \, e^{-ik_\perp\cdot b} f_{q/A}(x, k_\perp^2; Q),
\label{pdfs}
\\
D_{h/q}(z, b; Q) &= \frac{1}{z^2} \int d^2p_T \, e^{-ip_T\cdot b/z} D_{h/q}(z, p_T^2; Q),
\label{ffs}
\\
f_{1T}^{\perp q(\alpha)}(x, b; Q) &= \frac{1}{M} \int d^2k_\perp \, e^{-ik_\perp\cdot b} 
k_{\perp}^\alpha f_{1T}^{\perp q}(x, k_\perp^2; Q)
\,.
\label{sivers}
\end{align}
In this work we apply the Collins-Soper-Sterman (CSS) approach~\cite{Collins:1984kg,Qiu:2000ga,Landry:2002ix}, i.e., we choose $Q_i = c/b$ as the initial scale ($c=2e^{-\gamma_E}$) and perform the resummation at next-to-leading logarithmic (NLL) accuracy.

In the perturbative region $1/b\gg \Lambda_{\rm QCD}$ we can refactorize the TMD $F(x, b; \mu=c/b)$ in terms of the corresponding collinear function:
\bea
F_{i/h}(x, b; \mu) = 
\sum_a \int_x^1 \frac{d\xi}{\xi} 
C_{i/a}\left(\frac{x}{\xi},b; \mu \right) f_{a/h}(\xi, \mu)
+{\cal O}(b\,\Lambda_{\rm QCD})
\,,
\eea
where $C_{i/a}(z,b; \mu) = \sum_{n=0}^\infty C_{i/a}^{(n)}(\alpha_s/\pi)^n$ are the perturbative coefficients.
In order to extrapolate this result to the non-perturbative large-$b$ region we follow the standard CSS approach~\cite{Collins:1984kg,Landry:2002ix} and introduce a non-perturbative Sudakov factor $R_{NP}(x, b, Q)$:
\bea
F(x, b; Q) = F_{\rm pert}(x, b_{*}; Q) R_{NP}(x, b, Q) , 
\label{full-evolution}
\eea
where $b_{*} = b/\sqrt{1+(b/b_{\rm max})^2}$ and $b_{\rm max}$ is introduced such that 
$b_*\approx b$ at small $b\ll b_{\rm max}$ region, while it approaches the limit $b_{\rm max}$ 
when $b$ becomes non-perturbatively large. 

The non-perturbative Sudakov factor $R_{\rm NP}(b, Q) = \exp(-S_{\rm NP})$ has been extracted from experimental data and is mainly constrained by the large $Q$ fits~\cite{Landry:2002ix,Konychev:2005iy}. 
In our work we find a universal form such that can be used to describe the world data for SIDIS at relatively low $Q$, DY lepton pair production at intermediate $Q$ 
and $W/Z$ boson production at large $Q$.
It has the form~\cite{Landry:2002ix,Konychev:2005iy,Davies:1984sp,Ellis:1997sc}:
\bea
S_{\rm NP}^{\rm pdf}(b, Q) &= b^2\left(g_1^{\rm pdf}+ \frac{g_2}{2} \ln\frac{Q}{Q_0}\right),
\\
S_{\rm NP}^{\rm ff}(b, Q) &= b^2\left(g_1^{\rm ff}+ \frac{g_2}{2} \ln\frac{Q}{Q_0}\right),
\\
S_{\rm NP}^{\rm sivers}(b, Q) &=b^2\left(g_1^{\rm sivers} + \frac{g_2}{2} \ln\frac{Q}{Q_0}\right),
\eea
for the unpolarized TMDPDFs and TMDPFFs and the weighted quark Sivers function in Eqs.~\eqref{pdfs}, \eqref{ffs} and \eqref{sivers}. 
Notice that the parameter $g_2$ is universal.
On the other hand, the parameter $g_1$ depends on the type of TMD.
Assuming a Gaussian form we have
\bea
g_1^{\rm pdf} = \frac{\langle k_\perp^2\rangle_{Q_0}}{4},
\qquad
g_1^{\rm ff} = \frac{\langle p_T^2\rangle_{Q_0}}{4z^2},
\qquad
g_1^{\rm sivers} = \frac{\langle k_{s\perp}^2\rangle_{Q_0}}{4},
\eea
where $\langle k_\perp^2\rangle_{Q_0}$, $\langle p_T^2\rangle_{Q_0}$ and $\langle k_{s\perp}^2\rangle_{Q_0}$ 
are the averaged intrinsic transverse momenta squared at momentum scale $Q_0$.

\section{Unpolarized $q_T$-spectra}	

For single hadron production in SIDIS, $e(\ell) + A(P)\to e(\ell') + h(P_h) +X$,
where $A$ (also $B$ below) generically represents the incoming hadron and $h$ the observed hadron, we define the virtual photon momentum $q=\ell - \ell'$ and its invariant mass $Q^2 = - q^2$, and adopt the usual SIDIS variables \cite{Meng:1991da}:
\bea
S_{ep} = (P+\ell)^2,
\qquad
x_B = \frac{Q^2}{2P\cdot q},
\qquad
y=\frac{P\cdot q}{P\cdot \ell} = \frac{Q^2}{x_B S_{ep}},
\qquad
z_h = \frac{P\cdot P_h}{P\cdot q}.
\eea
The so-called hadron multiplicity distribution is defined as
\bea
\frac{dN}{dz_h d^2P_{h\perp}} = \left. \frac{d\sigma}{dx_B dQ^2 dz_h d^2P_{h\perp}}\right/ 
\frac{d\sigma}{dx_B dQ^2},
\label{hadron-mul}
\eea
where the numerator and denominator are given by
\begin{align}
\frac{d\sigma}{dx_B dQ^2 dz_h d^2P_{h\perp}} &= \frac{\sigma_0^{\rm DIS}}{2\pi} 
\sum_q e_q^2 \int_0^{\infty} db\, b J_0(P_{h\perp}b/z_h) f_{q/A}(x_B, b; Q) D_{h/q}(z_h, b; Q),
\label{sidis-pt}
\\
\frac{d\sigma}{dx_B dQ^2} &= \sigma_0^{\rm DIS} \sum_q e_q^2 f_{q/A}(x_B, Q),
\end{align}
with $\sigma_0^{\rm DIS} = 2\pi\alpha_{\rm em}^2\left[1+(1-y)^2\right]/Q^4$.
Notice that we have taken the hard factor at leading order (equal to 1) in the above TMD factorization formula and throughout the paper, to be consistent with the resummation order that we implement (NLL). 
Notice as well that the relevant soft function for each process is already included in the proper definition of the TMDs in each case~\cite{collins-book,Echevarria:2012js}.

For Drell-Yan lepton pair production, 
$A(P_A)+B(P_B)\to [\gamma^*\to] \ell^+\ell^-(y,Q, p_\perp) +X$, with $y, Q, p_\perp$ the rapidity, invariant mass and transverse momentum of the pair, respectively, the spin-averaged differential cross section is~\cite{GarciaEchevarria:2011rb}
\bea
\frac{d\sigma}{dQ^2 dy d^2p_\perp} = \frac{\sigma_0^{\rm DY}}{2\pi}
\sum_q e_q^2 \int_0^{\infty} db\, bJ_0(p_\perp b) f_{q/A}(x_a, b; Q) f_{\bar q/B}(x_b, b; Q)
\,,
\label{DY}
\eea
where $\sigma_0^{\rm DY} = 4\pi\alpha_{\rm em}^2/3sQ^2N_c$, $s=(P_A+P_B)^2$ is the center-of-mass (CM) energy squared, and the parton momentum fractions $x_a$ and $x_b$ are given by $x_a = \frac{Q}{\sqrt{s}} e^y$ and $x_b = \frac{Q}{\sqrt{s}} e^{-y}$.

Finally, for $W/Z$ production, $A(P_A)+B(P_B)\to W/Z(y, p_\perp)+X$, 
the differential cross sections are given by~\cite{Kang:2009bp,Kang:2009sm}
\bea
\frac{d\sigma^W}{dyd^2p_\perp} &= \frac{\sigma_0^W}{2\pi} \sum_{q,q'} 
|V_{qq'}|^2 \int_0^{\infty} db\, b J_0(q_\perp b) f_{q/A}(x_a, b; Q) f_{q'/B}(x_b, b; Q),
\label{w-cross}
\\
\frac{d\sigma^Z}{dyd^2p_\perp} &= \frac{\sigma_0^Z}{2\pi} \sum_{q} \left(V_q^2+A_q^2\right) 
\int_0^{\infty} db\, b J_0(q_\perp b) f_{q/A}(x_a, b; Q) f_{q'/B}(x_b, b; Q),
\label{z-cross}
\eea
where $V_{qq'}$ are the CKM matrix elements and $V_q$ and $A_q$ are 
the vector and axial couplings of the $Z$ boson to quarks, respectively. 
The LO cross sections $\sigma_0^W$ and $\sigma_0^Z$ are
\bea
\sigma_0^W = \frac{\sqrt{2} \pi G_F M_W^2}{sN_c},
\qquad
\sigma_0^Z = \frac{\sqrt{2} \pi G_F M_Z^2}{sN_c},
\eea
where $G_F$ is the Fermi weak coupling constant and $M_W$ ($M_Z$) is the mass of the $W$ ($Z$) boson.

In our work we find that, for $Q_0 = \sqrt{2.4}$ GeV and $b_{\rm max}=1.5$ GeV$^{-1}$, the world data for SIDIS at relatively low $Q$, DY lepton pair production at intermediate $Q$ and $W/Z$ boson production at large $Q$ can be reasonably well described with
\bea
\langle k_\perp^2\rangle_{Q_0} = 0.38 {\rm ~GeV}^2,
\qquad
\langle p_T^2\rangle_{Q_0} = 0.19 {\rm ~GeV}^2,
\qquad
g_2 = 0.16  {\rm ~GeV}^2
\,.
\label{sudakov}
\eea
We then use these parameters to perform a fit and extract Sivers function from SIDIS data.
Even though the description is just qualitatively good, we emphasize the fact that our formalism is the very first attempt to use a universal form to describe both SIDIS and DY data together, and that we perform the resummation at NLL accuracy.
A first attempt to implement the approach presented in~[\refcite{Echevarria:2012pw}] has been pursued in~[\refcite{D'Alesio:2014vja}].

\section{Extraction of Sivers asymmetry in SIDIS}

The SIDIS cross section for a transversely polarized nucleon target is~\cite{Kang:2012xf,Anselmino:2008sga}
\bea
\frac{d\sigma}{dx_B dy dz_h d^2P_{h\perp}}
&= \sigma_0(x_B, y, Q^2)
\left[F_{UU} +    \sin(\phi_h-\phi_s)\,
F_{UT}^{\sin\left(\phi_h -\phi_s\right)} \right],
  \label{eq:aut}
\eea
where $\sigma_0 = \frac{2\pi \alpha_{\rm em}^2}{x_B y\, Q^2}\left(1+(1-y)^2\right)$, 
and $\phi_s$ and $\phi_h$ are the azimuthal angles for the nucleon spin and the transverse momentum of the outgoing hadron, respectively.
$F_{UU}$ and $F_{UT}^{\sin(\phi_h-\phi_s)}$ are the spin-averaged and spin-dependent structure functions:
\begin{align}
F_{UU} &=  \frac{1}{2\pi} \int_0^\infty db\, b J_0(P_{h\perp} b/z_h)\sum_q e_q^2  
f_{q/A}(x_B, b; Q) D_{h/q}(z_h, b; Q),
\\
F_{UT}^{\sin\left(\phi_h -\phi_s\right)} & =  - \int\frac{d^2b}{(2\pi)^2} 
e^{-iP_{h\perp}\cdot b/z_h} \hat{P}_{h\perp}^{\alpha}\sum_q e_q^2  
f_{1T, \rm SIDIS}^{\perp q(\alpha)}(x_B, b; Q) D_{h/q}(z_h, b; Q),
\label{fut}
\end{align}
where $\hat{P}_{h\perp}$ is the unit vector along the hadron transverse momentum $P_{h\perp}$. 
The Sivers asymmetry $A_{UT}^{\sin(\phi_h-\phi_s)}$ is defined as
\bea
A_{UT}^{\sin(\phi_h-\phi_s)} = \frac{\sigma_0(x_B, y, Q^2)}{\sigma_0(x_B, y, Q^2)} 
\frac{F_{UT}^{\sin\left(\phi_h -\phi_s\right)}}{F_{UU}}
\,.
\eea

In order to fit Sivers asymmetry we parametrize the Qiu-Sterman function $T_{q,F}(x, x, \mu)$ following [\refcite{Kouvaris:2006zy}]:
\bea
T_{q, F}(x, x, \mu) = N_q \frac{(\alpha_q+\beta_q)^{(\alpha_q+\beta_q)}}{\alpha_q^{\alpha_q} \beta_q^{\beta^q}} 
x^{\alpha_q} (1-x)^{\beta_q} f_{q/A}(x, \mu). 
\eea
We will then have 11 fitting parameters in total: $\alpha_u, \alpha_d, N_u, N_d$ for $u$ and $d$ quarks; $N_{\bar u}, N_{\bar d}, N_{s}, N_{\bar s}, \alpha_{\rm sea}$ for sea quarks; $\beta_q\equiv \beta$ for all quark flavors; and 
$\langle k_{s\perp}^2\rangle = 4\,g_1^{\rm sivers}$.

We use the MINUIT package to perform the fit and restrict experimental data to:
hadron production at JLab~\cite{Qian:2011py} with $\langle Q^2\rangle = 1.38 - 2.68$ GeV$^2$ and $P_{h\perp}\leq 0.5$ GeV; hadron production at HERMES~\cite{Airapetian:2009ae} with $\langle Q^2\rangle \approx 2.45$ GeV$^2$ and $P_{h\perp}\leq 0.6$ GeV; and COMPASS experimental data~\cite{Alekseev:2008aa,Adolph:2012sp} with $\langle Q^2\rangle \approx 3 - 5$ GeV$^2$ and $P_{h\perp}\leq 0.7$ GeV.
We obtain a good overall description of the data with a total $\chi^2\approx 300$ for 241 data points, and thus $\chi^2/d.o.f. = 1.3$.
The value of the fitted parameters are shown in the Table.~\ref{fitpar}. 

\begin{table}[ht]
\tbl{Best values of the free parameters  for the Sivers function from our fit to 
SIDIS data~\protect\cite{Airapetian:2009ae,Alekseev:2008aa,Adolph:2012sp,Qian:2011py} on $A_{UT}^{\sin(\phi_h-\phi_s)}$.}
{\begin{tabular}{l l l l l l}
\hline
\hline
&&$\chi^2/d.o.f. = 1.3$&&&\\
\hline
$\alpha_{u}$ &=& $1.051^{+0.192}_{-0.180}$  & $\alpha_{d}$ &=&  $1.552^{+0.303}_{-0.275}$ \\
$\alpha_{\rm sea}$ &=& $0.851^{+0.307}_{-0.305}$ & $\beta$ &=& $4.857^{+1.534}_{-1.395} $ \\
$N_{u}$ &=& $0.106^{+0.011}_{-0.009}$ & $N_{d}$ &=& $-0.163^{+0.039}_{-0.046} $ \\
$N_{\bar u}$ &=&  $-0.012^{+0.018}_{-0.020}$ & $N_{\bar d}$ &=&   $-0.105^{+0.043}_{-0.060}$\\
$N_{s}$ &=& $0.103^{+0.548}_{-0.604}$ & $N_{\bar s}$ &=& $-1.000{\pm 1.757}$ \\
$\langle k_{s\perp}^2\rangle$ &=& $0.282^{+0.073}_{-0.066}$ GeV$^2$ & & & \\
\hline
\hline
\end{tabular} \label{fitpar}}
\end{table}

In Fig.~\ref{qiu-sterman-fit}  we show the Qiu-Sterman function $T_{q,F}(x, x, Q)$ extracted from our fit for $u$, $d$ and $s$ quarks. 
We find that the functions for $u$ and $d$ quarks have similar size but opposite sign~\cite{Anselmino:2008sga,Sun:2013hua}. 
The sea quark Sivers functions are not well constrained in the fit.
\bef
\psfig{file=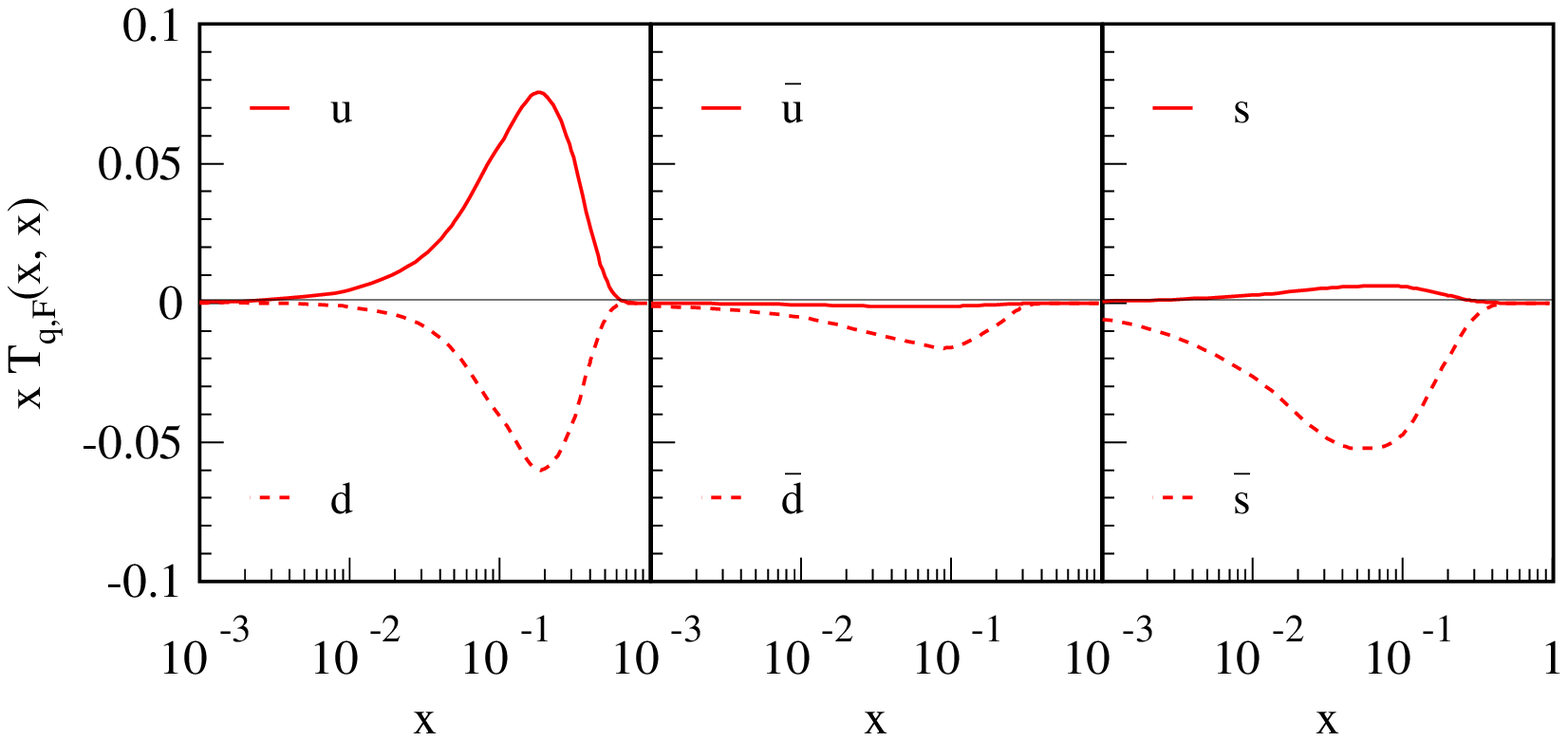, width=9cm}
\caption{The Qiu-Sterman function $T_{q,F}(x, x, Q)$ for $u$, $d$, and $s$ flavors at a 
scale $Q^2=2.4$ GeV$^2$, as extracted by our simultaneous fit of JLab, HERMES, and COMPASS data.}
\label{qiu-sterman-fit}
\eef

\section{Prediction of Sivers asymmetry in DY}

For Drell-Yan production, 
$A^\uparrow(P_A, s_\perp) + B(P_B) \to [\gamma^*\to] \ell^+\ell^-(Q, q_\perp)+X$, the spin-dependent cross section $\Delta \sigma \equiv \left[\sigma(s_\perp)-\sigma(-s_\perp)\right]/2$ can be written 
as~\cite{Kang:2011mr,Kang:2012vm}
\begin{align}
\frac{d\Delta\sigma}{dQ^2 dy d^2p_\perp} &= \epsilon^{\alpha\beta} s_\perp^\alpha 
\sigma_0^{\rm DY} \int \frac{d^2b}{(2\pi)^2} e^{-ip_\perp\cdot b} \sum_q e_q^2\, 
f_{1T, \rm DY}^{\perp,q(\beta)}(x_a, b; Q)  f_{\bar q/B}(x_b, b; Q),
\nnu
&= - \frac{\sigma_0^{\rm DY}}{4\pi} \int_0^{\infty} db\, b^2J_1(p_\perp b) 
\sum \sum_q e_q^2\, T_{q, F}(x_a, x_a, c/b^*) 
f_{\bar q/B}(x_b, c/b^*)  
\nnu
&\times
\exp\left\{-\int_{c/b^*}^Q \frac{d\mu^2}{\mu^2} \left(A\ln\frac{Q^2}{\mu^2}+B\right)\right\} 
\exp\left\{-S_{\rm NP}^{\rm sivers}(b, Q)\right\}.
\label{spin-dy}
\end{align}
To obtain the second expression in the above equation we apply the sign change 
for the Sivers functions between SIDIS and DY processes:
\bea
f_{1T, \rm DY}^{\perp,q(\beta)}(x_a, b; Q)  = - f_{1T, \rm SIDIS}^{\perp,q(\beta)}(x_a, b; Q).
\eea
The single transverse spin asymmetry for DY production is then given by
\bea
A_N = \left.\frac{d\Delta\sigma}{dQ^2 dy d^2p_\perp}\right/ \frac{d\sigma}{dQ^2 dy d^2p_\perp}.
\eea

Several experiments are planned to measure the Sivers asymmetry in DY production: COMPASS collaboration~\cite{DY-compass-exp}, Fermilab~\cite{DY-fermi-beam,DY-fermi-target} and RHIC~\cite{Aschenauer:2013woa,DY-rhic-exp}.
Below we present our predictions for $A_N$ based on our fit.

\bef
\psfig{file=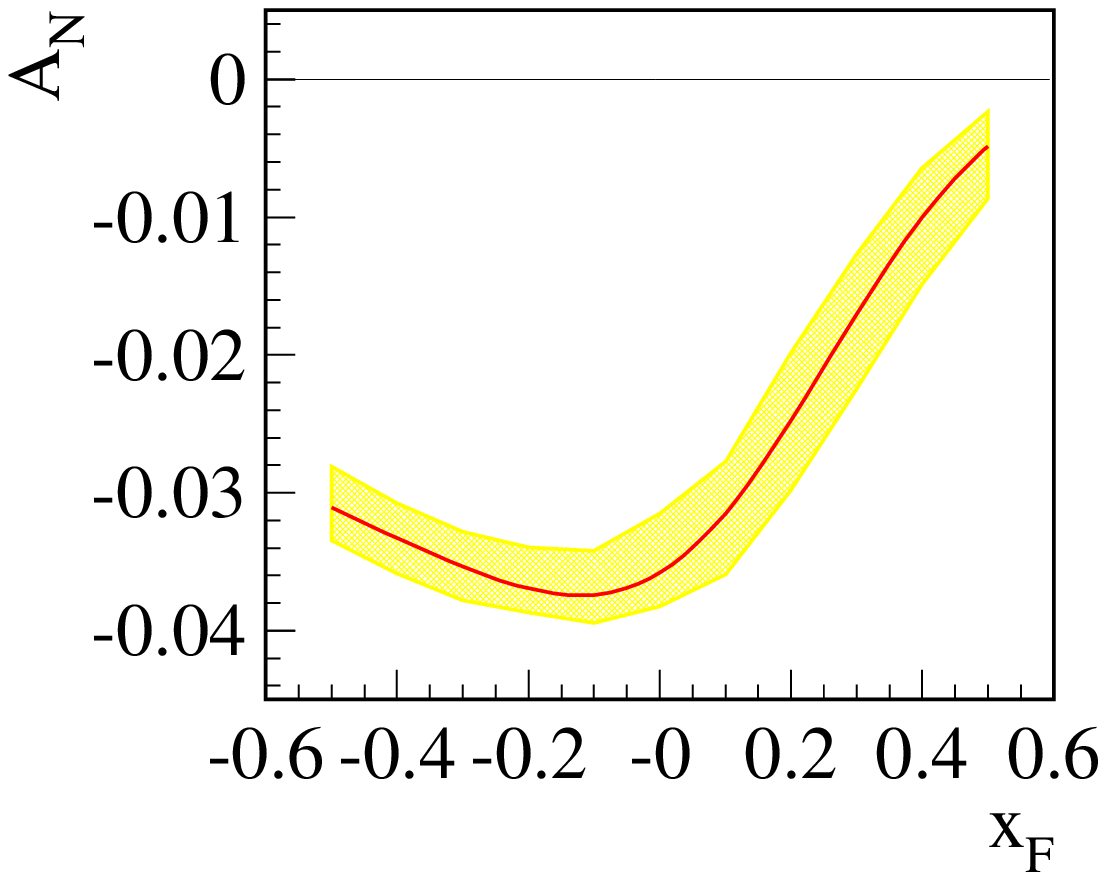, width=3.8cm}
\quad
\psfig{file=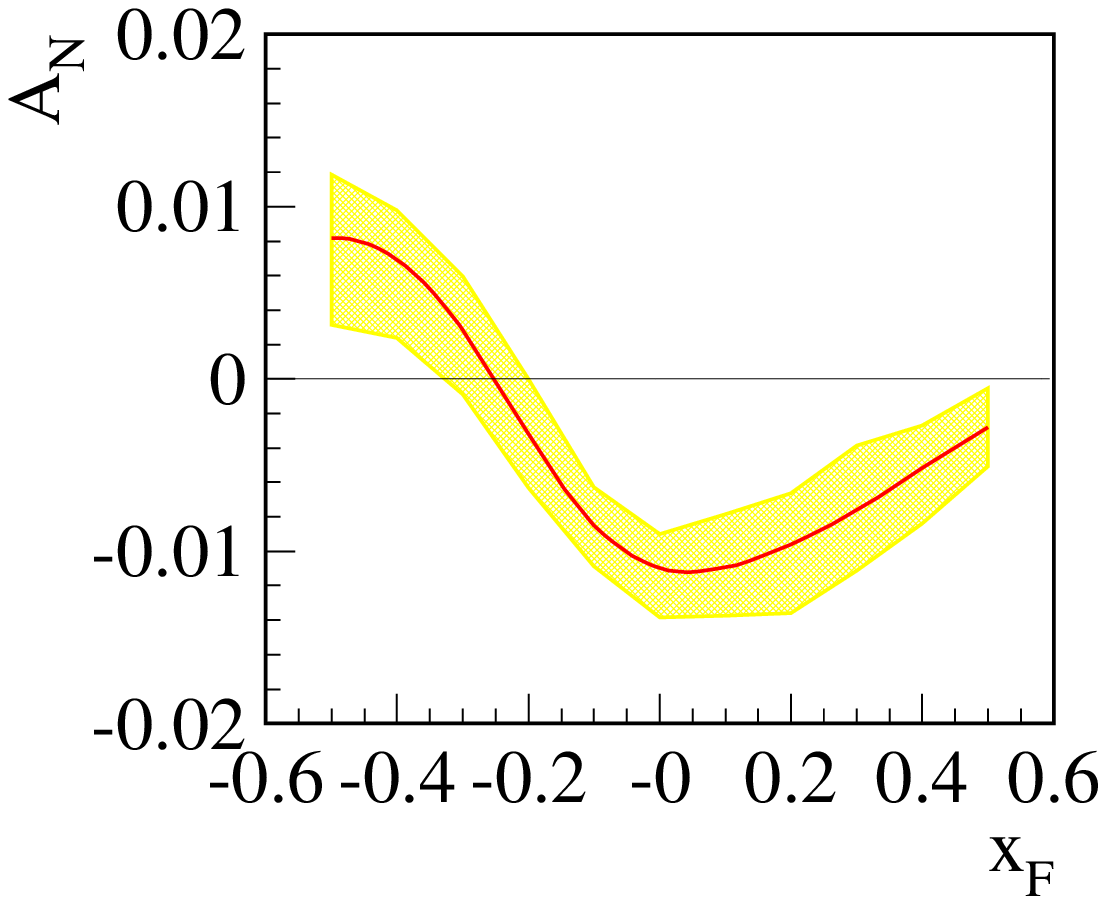, width=3.8cm}
\quad
\psfig{file=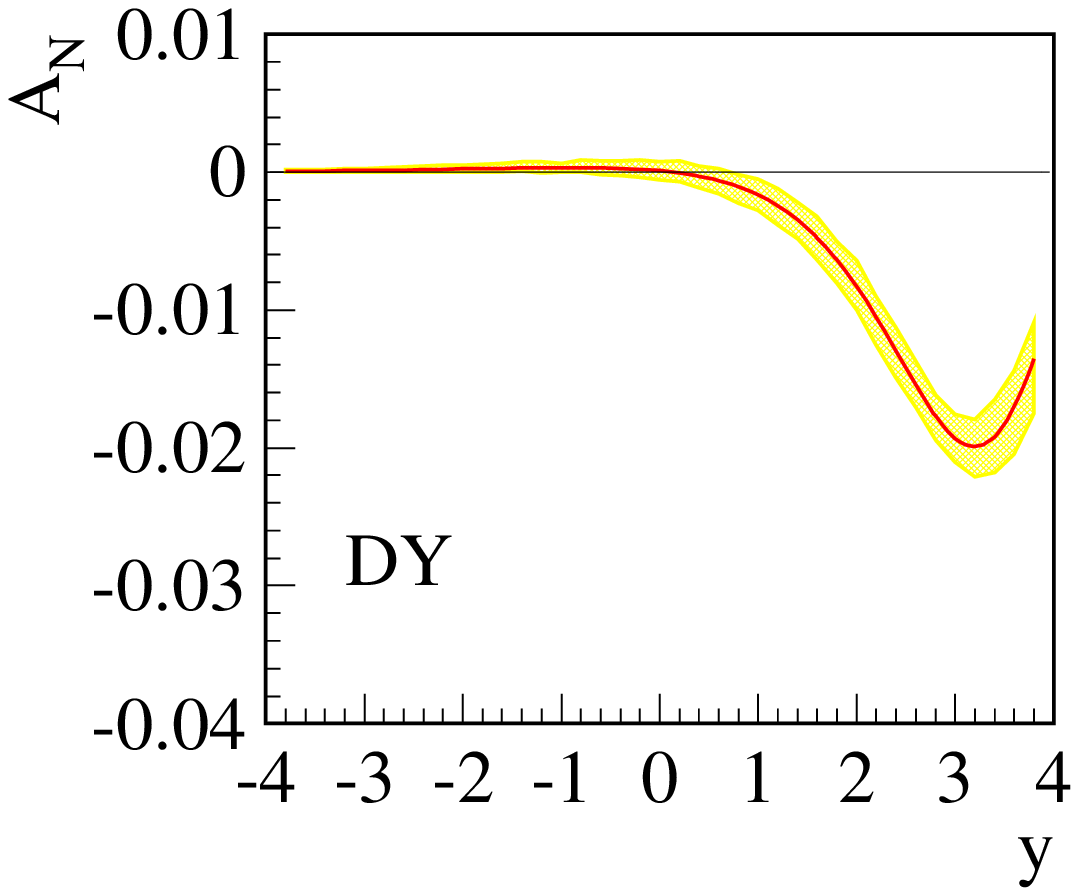, width=3.8cm}
\caption{The estimated Sivers asymmetries for DY lepton pair production. Left plot: $A_N$ in $p^\uparrow \pi^-$ collisions 
as a function of $x_F$ at COMPASS energy $\sqrt{s}=18.9$ GeV. Middle plot: $A_N$ in $p^\uparrow p$ collisions is plotted as a 
function of $x_F$ at Fermilab energy $\sqrt{s}=15.1$ GeV. Right plot: $A_N$ in $p^\uparrow p$ collisions is plotted 
as a function of the pair's rapidity $y$ at RHIC energy $\sqrt{s}=510$ GeV. We have integrated over
the pair's transverse momentum $0<p_\perp<1$ GeV in the invariant mass range $4<Q<9$ GeV.}
\label{DY-sivers}
\eef

\bef
\psfig{file=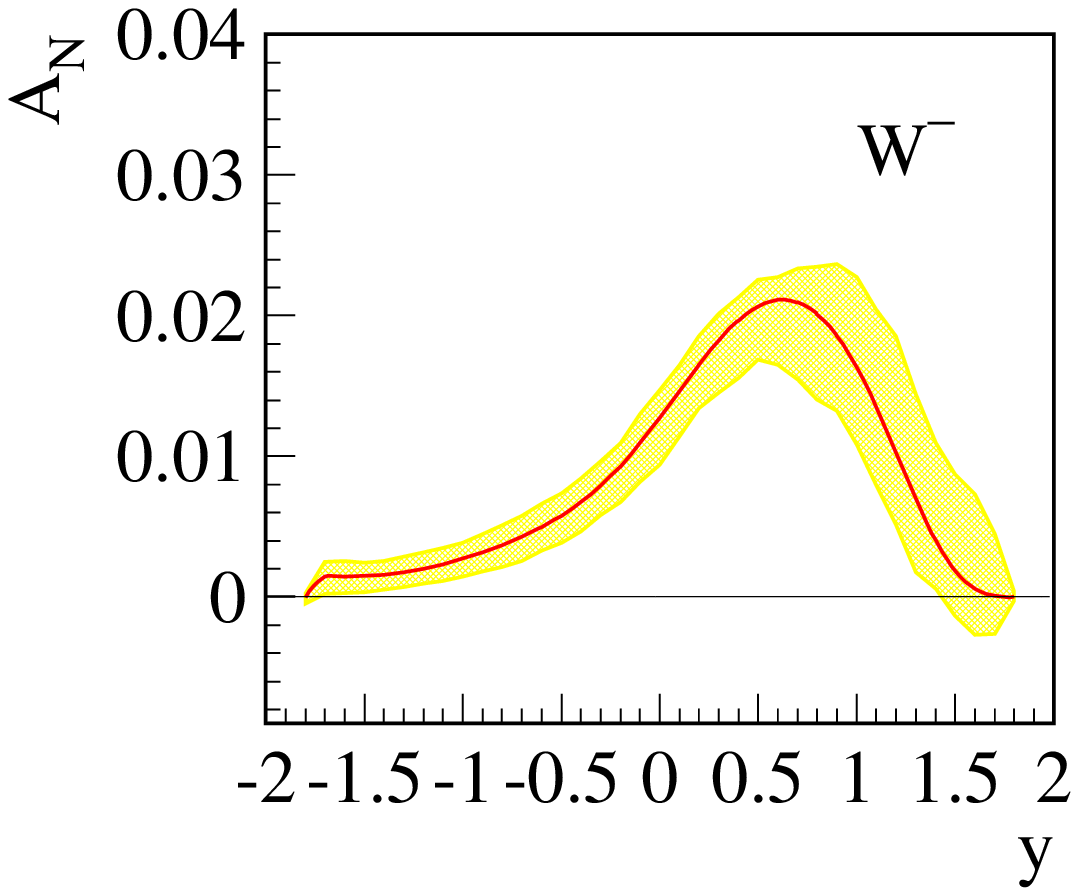, width=3.8cm}
\quad
\psfig{file=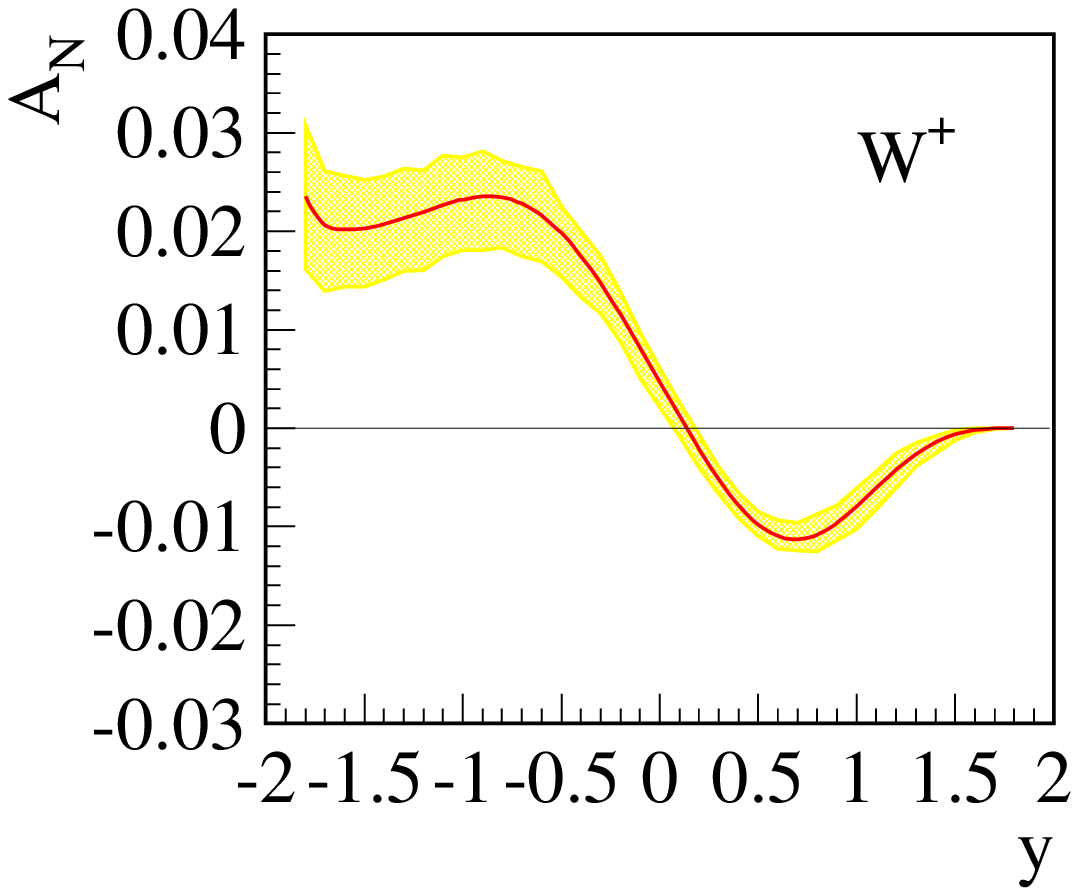, width=3.8cm}
\caption{The estimated Sivers asymmetries as a function of rapidity $y$ for $W^-$  and $W^+$ production at the  
RHIC energy $\sqrt{s}=510$ GeV. We have integrated over the 
transverse momentum for $W$ boson in $0<p_\perp <3$ GeV.}
\label{Wboson}
\eef

In Fig.~\ref{DY-sivers} (left) we plot our prediction for Sivers asymmetry $A_N$ for DY lepton pair production as a function of $x_F=x_a-x_b$ for COMPASS kinematics. 
The solid curve is obtained with the parameters in Table.~\ref{fitpar}, while the band comes from their 1$\sigma$ error.
Since sea quark Sivers functions are not really constrained by the current experimental data, thus if one includes the full uncertainty from the sea quarks the band could be much larger in the region where the sea quarks dominate (e.g., the negative $x_F$ or $y$ regions). 
COMPASS projected their measurement around $x_F=-0.2$, where the estimated asymmetry we obtain is around $3-4\%$, and thus measurable. 
In Fig.~\ref{DY-sivers} (middle) we show the predicted Sivers asymmetry for the Fermilab kinematics. 
The proposed ``polarized beam'' experiment~\cite{DY-fermi-beam} will correspond to $0<x_F<0.6$, while the proposed ``polarized target'' experiment~\cite{DY-fermi-target} will correspond to $-0.6<x_F<0.1$.
The asymmetry we obtain is around $1-2\%$, which we hope could be measured.
Finally, in Fig.~\ref{DY-sivers} (right) we show $A_N$ as a function of the lepton pair's rapidity $y$ at RHIC kinematics. 
We find an asymmetry of $2-3\%$ in the forward rapidity region, which should be measurable. 

$W$ boson asymmetries have also been proposed~\cite{Kang:2009bp} to measure the Sivers asymmetry, and have been planned at RHIC experiment~\cite{Aschenauer:2013woa}. 
In Fig.~\ref{Wboson} we show our prediction for $A_N$ as a function of the rapidity $y$ for $W^-$ and $W^+$ boson production. 
The asymmetry is about $2 - 3\%$, which we hope can still be measured by the 
RHIC experiment.

\section{Conclusions}
We have performed a phenomenological extraction of Sivers asymmetry from current SIDIS data, paying special attention to the QCD evolution of the TMDs involved and the non-perturbative Sudakov factors.
Since one important part of those is spin-independent, we have first found a universal form that allows us to describe reasonably well the world's data for the transverse momentum distribution in SIDIS at relatively low momentum scale $Q$, DY lepton pair production at intermediate $Q$, and $W/Z$ production at high $Q$.
Then we performed the fit of Sivers asymmetry using SIDIS data from HERMES, COMPASS and Jefferson Lab, and use it to make predictions for DY, which should be compared to future experimental data in DY production to test its sign change and better constrain the Sivers function for sea quarks.

\section*{Acknowledgements}
This research is supported by the US Department of Energy, Office of Science, by the LDRD program at LANL, the research program of the ``Stichting voor Fundamenteel Onderzoek der Materie (FOM)'', which is financially supported by the ``Nederlandse Organisatie voor Wetenschappelijk Onderzoek (NWO)'', and the US Department of Energy under grant number DE-SC0008745.



\begin{thebibliography}{99}

 \bibitem{collins-book}
J.~C.~Collins, {\it Foundations of Perturbative QCD} (Cambridge University Press, Camberg, 2011).

\bibitem{Echevarria:2012js}
  M.~G.~Echevarria, A.~Idilbi and I.~Scimemi,
  Phys.\ Lett.\ B {\bf 726} (2013) 795
  [arXiv:1211.1947 [hep-ph]].


\bibitem{Sivers:1989cc}
  D.~W.~Sivers,
  Phys.\ Rev.\ D {\bf 41} (1990) 83.


\bibitem{Brodsky:2002cx}
  S.~J.~Brodsky, D.~S.~Hwang and I.~Schmidt,
  Phys.\ Lett.\ B {\bf 530} (2002) 99
  [hep-ph/0201296].


\bibitem{Collins:2002kn}
  J.~C.~Collins,
  Phys.\ Lett.\ B {\bf 536} (2002) 43
  [hep-ph/0204004].


\bibitem{Boer:2003cm}
  D.~Boer, P.~J.~Mulders and F.~Pijlman,
  Nucl.\ Phys.\ B {\bf 667} (2003) 201
  [hep-ph/0303034].


\bibitem{Kang:2011hk}
  Z.~B.~Kang, J.~W.~Qiu, W.~Vogelsang and F.~Yuan,
  Phys.\ Rev.\ D {\bf 83} (2011) 094001
  [arXiv:1103.1591 [hep-ph]].


\bibitem{Kang:2009bp}
  Z.~B.~Kang and J.~W.~Qiu,
  Phys.\ Rev.\ Lett.\  {\bf 103} (2009) 172001
  [arXiv:0903.3629 [hep-ph]].


\bibitem{Echevarria:2014xaa}
  M.~G.~Echevarria, A.~Idilbi, Z.~B.~Kang and I.~Vitev,
  Phys.\ Rev.\ D {\bf 89} (2014) 074013
  [arXiv:1401.5078 [hep-ph]].


\bibitem{Aybat:2011zv}
  S.~M.~Aybat and T.~C.~Rogers,
  Phys.\ Rev.\ D {\bf 83} (2011) 114042
  [arXiv:1101.5057 [hep-ph]].


\bibitem{Aybat:2011ge}
  S.~M.~Aybat, J.~C.~Collins, J.~W.~Qiu and T.~C.~Rogers,
  Phys.\ Rev.\ D {\bf 85} (2012) 034043
  [arXiv:1110.6428 [hep-ph]].


\bibitem{Echevarria:2012pw}
  M.~G.~Echevarria, A.~Idilbi, A.~Schafer and I.~Scimemi,
  Eur.\ Phys.\ J.\ C {\bf 73} (2013) 2636
  [arXiv:1208.1281 [hep-ph]].


\bibitem{Collins:1984kg}
  J.~C.~Collins, D.~E.~Soper and G.~F.~Sterman,
  Nucl.\ Phys.\ B {\bf 250} (1985) 199.


\bibitem{Qiu:2000ga}
  J.~w.~Qiu and X.~f.~Zhang,
  Phys.\ Rev.\ Lett.\  {\bf 86} (2001) 2724
  [hep-ph/0012058].


\bibitem{Landry:2002ix}
  F.~Landry, R.~Brock, P.~M.~Nadolsky and C.~P.~Yuan,
  Phys.\ Rev.\ D {\bf 67} (2003) 073016
  [hep-ph/0212159].


\bibitem{Konychev:2005iy}
  A.~V.~Konychev and P.~M.~Nadolsky,
  Phys.\ Lett.\ B {\bf 633} (2006) 710
  [hep-ph/0506225].


\bibitem{Davies:1984sp}
  C.~T.~H.~Davies, B.~R.~Webber and W.~J.~Stirling,
  Nucl.\ Phys.\ B {\bf 256} (1985) 413.


\bibitem{Ellis:1997sc}
  R.~K.~Ellis, D.~A.~Ross and S.~Veseli,
  Nucl.\ Phys.\ B {\bf 503} (1997) 309
  [hep-ph/9704239].


\bibitem{Meng:1991da}
  R.~b.~Meng, F.~I.~Olness and D.~E.~Soper,
  Nucl.\ Phys.\ B {\bf 371} (1992) 79.


\bibitem{GarciaEchevarria:2011rb}
  M.~G.~Echevarria, A.~Idilbi and I.~Scimemi,
  JHEP {\bf 1207} (2012) 002
  [arXiv:1111.4996 [hep-ph]].


\bibitem{Kang:2009sm}
  Z.~B.~Kang and J.~W.~Qiu,
  Phys.\ Rev.\ D {\bf 81} (2010) 054020
  [arXiv:0912.1319 [hep-ph]].
  
\bibitem{D'Alesio:2014vja}
  U.~D'Alesio, M.~G.~Echevarria, S.~Melis and I.~Scimemi,
  arXiv:1407.3311 [hep-ph].


\bibitem{Kang:2012xf}
  Z.~B.~Kang and A.~Prokudin,
  Phys.\ Rev.\ D {\bf 85} (2012) 074008
  [arXiv:1201.5427 [hep-ph]].


\bibitem{Anselmino:2008sga}
  M.~Anselmino, M.~Boglione, U.~D'Alesio, A.~Kotzinian, S.~Melis, F.~Murgia, A.~Prokudin and C.~Turk,
  Eur.\ Phys.\ J.\ A {\bf 39} (2009) 89
  [arXiv:0805.2677 [hep-ph]].

\bibitem{Kouvaris:2006zy}
  C.~Kouvaris, J.~W.~Qiu, W.~Vogelsang and F.~Yuan,
  Phys.\ Rev.\ D {\bf 74} (2006) 114013
  [hep-ph/0609238].

\bibitem{Qian:2011py}
  X.~Qian {\it et al.}  [Jefferson Lab Hall A Collaboration],
  Phys.\ Rev.\ Lett.\  {\bf 107} (2011) 072003
  [arXiv:1106.0363 [nucl-ex]].


\bibitem{Airapetian:2009ae}
  A.~Airapetian {\it et al.}  [HERMES Collaboration],
  Phys.\ Rev.\ Lett.\  {\bf 103} (2009) 152002
  [arXiv:0906.3918 [hep-ex]].


\bibitem{Alekseev:2008aa}
  M.~Alekseev {\it et al.}  [COMPASS Collaboration],
  Phys.\ Lett.\ B {\bf 673} (2009) 127
  [arXiv:0802.2160 [hep-ex]].


\bibitem{Adolph:2012sp}
  C.~Adolph {\it et al.}  [COMPASS Collaboration],
  Phys.\ Lett.\ B {\bf 717} (2012) 383
  [arXiv:1205.5122 [hep-ex]].


\bibitem{Sun:2013hua}
  P.~Sun and F.~Yuan,
  Phys.\ Rev.\ D {\bf 88} (2013) 11,  114012
  [arXiv:1308.5003 [hep-ph]].


\bibitem{Kang:2011mr}
  Z.~B.~Kang, B.~W.~Xiao and F.~Yuan,
  Phys.\ Rev.\ Lett.\  {\bf 107} (2011) 152002
  [arXiv:1106.0266 [hep-ph]].


\bibitem{Kang:2012vm}
  Z.~B.~Kang and B.~W.~Xiao,
  Phys.\ Rev.\ D {\bf 87} (2013) 034038
  [arXiv:1212.4809 [hep-ph]].

\bibitem{DY-compass-exp}
COMPASS proposal at CERN, \\
\url{http://wwwcompass.cern.ch/compass/proposal/compass-II_proposal/compass-II_proposal.pdf}.

\bibitem{DY-fermi-beam}
Fermilab DY proposal - polarized target, \\
\url{http://www.fnal.gov/directorate/program_planning/June2012Public/P-1027_Pol-Drell-Yan-proposal.pdf}. 
 
\bibitem{DY-fermi-target} 
Fermilab DY proposal - polarized beam, \\
\url{http://www.fnal.gov/directorate/program_planning/June2013PACPublic/P-1039_LOI_polarized_DY.pdf}.



\bibitem{Aschenauer:2013woa}
  E.~C.~Aschenauer, A.~Bazilevsky, K.~Boyle, K.~O.~Eyser, R.~Fatemi, C.~Gagliardi, M.~Grosse-Perdekamp and J.~Lajoie {\it et al.},
  arXiv:1304.0079 [nucl-ex].


\bibitem{DY-rhic-exp}
A$_{\rm N}$DY proposal, \url{http://www.bnl.gov/npp/docs/pac0611/DY_pro_110516_final.2.pdf}.
\end{thebibliography}
\end{document}